\documentclass[twocolumn,showpacs,prl]{revtex4}

\usepackage{graphicx}
\usepackage{dcolumn}
\usepackage{bm}
\usepackage{xcolor,amsmath}

\newcommand{\ket}[1]{|#1\rangle}
\newcommand{\bra}[1]{\langle#1|}

\newcommand{\eq}{\begin{equation}}
\newcommand{\fine}{\end{equation}}
\newcommand{\el}{\ell}
\newcommand{\cluster}{\ket{\Phi^{\text{lin}}_4}}
\begin{document}


\title{
One-way quantum computation via manipulation of polarization and momentum qubits in two-photon cluster states
}
\author{Giuseppe Vallone}
\homepage{http://quantumoptics.phys.uniroma1.it/}
\author{Enrico Pomarico}
\homepage{http://quantumoptics.phys.uniroma1.it/}
\author{Francesco De Martini}
\homepage{http://quantumoptics.phys.uniroma1.it/}
\author{Paolo Mataloni}
\homepage{http://quantumoptics.phys.uniroma1.it/}
\affiliation{
Dipartimento di Fisica dell'Universit\'{a} ``La Sapienza'' and
Consorzio Nazionale Interuniversitario per le Scienze Fisiche della Materia,
Roma, 00185 Italy}

\date{\today}

\begin{abstract}
Four-qubit cluster states of two photons entangled in polarization and linear momentum have been used to realize
a complete set of single qubit rotations and the C-NOT gate for equatorial qubits with high values of fidelity. By the computational
equivalence of the two degrees of freedom our result demonstrate the suitability of two photon cluster states for rapid and
efficient one-way quantum computing. 
\end{abstract}

\maketitle

The relevance of cluster states in quantum information and quantum computation (QC) has been emphasized 
in several papers in recent years \cite{01-bri-aon,05-kie-exp,05-wal-exp,07-pre-hig}.
By these states novel significant tests of quantum nonlocality, 
which are more resistant to noise and show significantly larger deviations from classical bounds can be realized
\cite{05-cab-str,05-sca-non,07-val-rea,05-kie-exp}.  
Besides that, cluster states represent today the basic resource for the realization of a quantum computer operating in the one-way model \cite{01-bri-aon}. 
In the standard QC approach any quantum algorithm can be realized by a sequence of single qubit rotations and two qubit gates, 
such as C-NOT and C-Phase \cite{01-kni-asc}. A deterministic one-way QC is based on the initial preparation of entangled qubits 
in a cluster state, followed by a temporally ordered pattern of single qubit measurements and feed-forward operations.
Indeed these operations, depending on the outcome of the measured qubits \cite{01-bri-aon}, 
correspond either to intermediate feed-forward measurements or to Pauli matrix feed-forward corrections on the final output state.
Two qubit gates can be realized by exploiting the existing entanglement between qubits.
In this way the difficulties of standard QC,
related to the implementation of two qubit gates, are transferred to the preparation of the state.
  
The first experimental results of one-way QC were demonstrated by using 4-photon cluster states
\cite{05-wal-exp,07-pre-hig}. 
The detection rate in such experiments, approximately 1 Hz, is limited by the fact that four photon events in a 
standard spontaneous parametric down conversion (SPDC) process are rare. 
Moreover, four-photon cluster states are characterized by limited values of fidelity, while 
efficient computation require highly faithful prepared states.

Cluster states at high level of brightness and fidelity were realized by entangling 
two photons in more degrees of freedom \cite{07-val-rea}. 
Precisely, this was demonstrated by entangling the polarization ($\pi$) and linear momentum ({\bf k})
degrees of freedom of one of the two photons belonging to a hyperentangled state \cite{05-bar-pol,06-bar-enh}. 
The high fidelity and detection rate of the prepared two-photon four-qubit cluster states 
make them suitable for the realization of high speed one-way QC.
\begin{figure}[t]
	\begin{center}
		\includegraphics[scale=0.3]{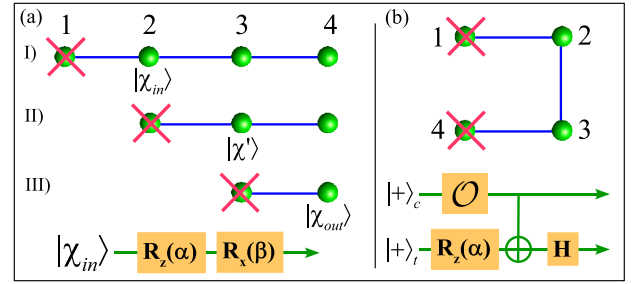}
	\end{center}
		\caption{Measurement pattern for single qubit rotations (a) and C-NOT gate (b). (a) Top: arbitrary single qubit rotations on a four 
		qubit linear cluster state are carried out in three steps (I, II, III). In each measurement, indicated by a red cross, the information 
		travels from left to right. Bottom: equivalent logical circuit. (b) C-NOT gate realization 
		via measurement of qubits 1, 4 on the horseshoe cluster (top) and equivalent circuit (bottom).}
\label{fig:cluster}
\end{figure}
In this letter we present the realization of arbitrary single 
qubit rotations and of the C-NOT gate for equatorial qubits. 
We have also verified the equivalence existing 
between the two degrees of freedom for qubit rotations, by using either 
$\bf k$ or $\pi$ as output qubit, demonstrating that all four qubits can be adopted for computational applications. 

In our experiment two-photon four-qubit cluster states are generated by starting from $\pi$-$\bf k$ 
hyperentangled photon pairs generated by SPDC process in continuous way. 
The method to produce the hyperentangled states 
$\ket{\Xi^{\pm\pm}}= \ket{\Phi^\pm}_\pi\otimes\ket{{\psi^\pm}}_{\bf k}$ was explained  in other papers 
\cite{05-bar-pol,05-cin-all}, to which we refer for details. 
In the above equations
$|\Phi^\pm\rangle_\pi =\frac{1}{\sqrt{2}}
\left( |H\rangle _{A}|H\rangle_{B}\pm|V\rangle _{A}|V\rangle _{B}\right)$,
$|{\psi^\pm} \rangle_{\bf k} =\frac{1}{\sqrt{2}}(|\ell \rangle_{A}|r\rangle _{B}\pm|r\rangle _{A}|\ell \rangle _{B})$, 
$H,V$ correspond to the horizontal ($H$) and
vertical ($V$) polarizations and $\ell ,r$ refer to the left ($\ell $) or
right ($r$) paths of the photon $A$ (Alice) or $B$ (Bob) (see Fig. 1 of \cite{07-val-rea}).

The cluster state
\eq
\begin{aligned}\label{cluster}
\ket{C_{4}}=&\frac{1}{2}(|H\ell \rangle _{A}|Hr\rangle _{B}- |Hr\rangle _{A}|H\ell \rangle_{B}\\
&+|Vr\rangle _{A}|V\ell \rangle _{B}+|V\ell \rangle _{A}|Vr\rangle_{B})
\end{aligned}
\fine
is created by starting from the state $\ket{\Xi^{+-}}=\ket{\Phi^+}_{\pi}\otimes\ket{\psi^-}_{\bf k}$ 
and introducing a $\pi$ phase shift in one of the four output modes of the SPDC source
\footnote{The state \eqref{cluster} is equivalent to that generated in \cite{07-val-rea} up to single qubit transformations.}. 
Precisely, a zero-order half wave (HW) plate is inserted on the $r_A$ mode. 
This operation, corresponding to a Controlled Phase (CP) between the control ($\bf k_A$) and the target ($\pi_A$) qubits,
create a genuine four-partite entanglement, without any kind of post-selection.
Cluster states were observed at 1 kHz detection rate with
fidelity $F = 0.880\pm0.013$, 
obtained from the measurement of the stabilizer operators of $\ket{C_4}$ \cite{05-kie-exp}.   

By using the correspondence $\ket H\leftrightarrow\ket0$, $\ket V\leftrightarrow\ket1$,
$\ket \el\leftrightarrow\ket0$, $\ket r\leftrightarrow\ket1$,
the generated state $\ket{C_4}$  is equivalent to the cluster state 
$\cluster=\frac12(\ket{+}_1\ket{0}_2\ket{0}_3\ket{+}_4+\ket{+}_1\ket{0}_2\ket{1}_3\ket{-}_4+
\ket{-}_1\ket{1}_2\ket{0}_3\ket{+}_4-\ket{-}_1\ket{1}_2\ket{1}_3\ket{-}_4)$ 
\footnote{The state $\cluster$ is obtained by preparing a chain of qubits all prepared in the state $\ket+$ and then
applying the gate $CP=\ket0\bra0\otimes\openone+\ket1\bra1\otimes\sigma_z$ for each link.}
(with $\ket{\pm}=\frac{1}{\sqrt2}(\ket0\pm\ket1)$)
up to single qubit unitaries:
\eq\label{equiv cluster}
\ket{C_4}=U_1\otimes U_2\otimes U_3\otimes U_4\cluster\equiv\mathcal U\cluster\,.
\fine
Here $\cluster$ and $\ket{C_4}$ are respectively expressed in the so called ``computational'' and 
``laboratory'' basis, while the $U_j$'s ($j=1,\cdots,4$) are products of Hadamard gates 
$H=\frac1{\sqrt2}(\sigma_x +\sigma_z)$ and Pauli matrices $\sigma_i$. 
Their explicit expressions depend on the ordering of the four physical qubits,
namely ${\bf k}_A$, ${\bf k}_B$, $\pi_A$, $\pi_B$.
In this work we use three different ordering:
\eq\notag
\begin{aligned}
a)& \text{(1,2,3,4)=}({\bf k}_B,{\bf k}_A,\pi_A,\pi_B), \mathcal U=\sigma_xH\otimes\sigma_z\otimes\openone\otimes H\\
b)& \text{(1,2,3,4)=}(\pi_B,\pi_A,{\bf k}_A,{\bf k}_B), \mathcal U=H\otimes\sigma_z\otimes\sigma_x\otimes \sigma_zH\\
c)& \text{(1,2,3,4)=}({\bf k}_A,{\bf k}_B,\pi_B,\pi_A), \mathcal U=\sigma_zH\otimes\sigma_x\otimes\openone\otimes H.
\end{aligned}
\fine
In the following we refer to these expressions depending on the logical operation we will consider.
Let's now examine the most relevant measurements performed in this experiment.

{\bf Single qubit rotations}.
In the one-way model a three-qubit linear cluster state is sufficient to realize
arbitrary single qubit rotations \cite{03-rau-mea,05-tam-qua}, 
hence this operation may be implemented through three different steps and 
single qubit measurements. According to the measurement basis for a generic qubit $j$, $\ket{\varphi_{\pm}}_j=\frac{1}{\sqrt2}(\ket0_j\pm e^{-i\varphi}\ket1_j)$,
we define $s_j=0$ ($s_j=1$) when the $\ket{\varphi_+}_j$ ($\ket{\varphi_-}_j$) outcome is obtained. 
With the four-qubit cluster expressed in the computational basis the following procedure must be performed (see fig. \ref{fig:cluster}(a)):

I: A three-qubit linear cluster is generated by measuring the first qubit in the basis $\{\ket{0}_1, \ket1_1\}$. 
The input logical qubit $\ket{\chi_{in}}$ is then encoded in qubit 2. If the outcome of the first measurement is 
$\ket0_1$ then $\ket{\chi_{in}} = \ket+$, otherwise $\ket{\chi_{in}}=\ket-$.

II: Measuring qubit 2 in the basis $\ket{\alpha_\pm}_2$, with $\alpha$ corresponding to a particular 
value of $\varphi$, the computational qubit (now encoded in qubit 3) 
is transformed into
$\ket{\chi'}={(\sigma_x)}^{s_2}HR_z(\alpha)\ket{\chi_{in}}$,
with $R_z(\alpha)=e^{-\frac{i}{2}\alpha \sigma_z}$.

III: Measurement of qubit 3 in the basis $\ket{\beta_\pm}_3$ (if $s_2=0$) or $\ket{-\beta_\pm}_3$ 
(if $s_2=1$) leaves the last qubit in the state 
$\ket{\chi_{out}}={(\sigma_x)}^{s_3}HR_z\left[(-1)^{s_2}\beta\right]\ket{\chi'}
=\sigma^{s_3}_x\sigma^{s_2}_zR_x(\beta)R_z(\alpha)\ket{\chi_{in}}$,
with $R_x(\beta)=e^{-\frac{i}{2}\beta \sigma_x}$.

In this way, by suitable choosing the values of $\alpha$ and $\beta$,
we can perform any arbitrary single qubit rotation $\ket{\chi_{in}}\rightarrow
R_x(\beta)R_z(\alpha)\ket{\chi_{in}}$
\footnote{Three sequential rotations are necessary to implement a generic $SU(2)$ matrix but only two, 
namely $R_x(\beta)R_z(\alpha)$, are sufficient to rotate the input state $\ket{\chi_{in}}=\ket\pm$ into
a generic state}
up to Pauli errors ($\sigma^{s_3}_x\sigma^{s_2}_z$), that can be corrected by proper feed-forward 
operations \cite{07-pre-hig}. 
\begin{figure}
	\centering
	\includegraphics[scale=.5]{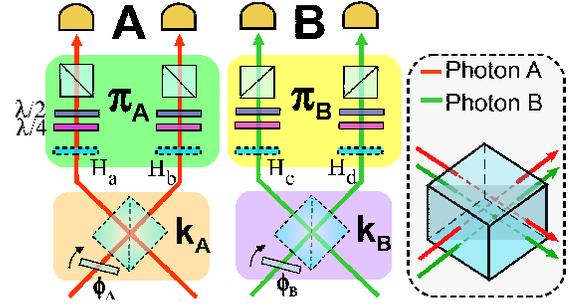}
	\caption{Measurement setup for photons A and B. Momentum qubits $\bf{k}_A$ and $\bf{k}_B$ are measured by 
phase shifters ($\phi_A$, $\phi_B$) and a common 50:50 BS. Polarization qubits $\pi_A$ and $\pi_B$ are measured
by $\lambda/4$, $\lambda/2$ waveplate and polarizing beam splitters PBS's. 
Hadamard gates $H_{a,b,c,d}$ are realized by HW plates. 
Dashed lines for $H_{a,b,c,d}$ and $BS$ indicate that these devices can be inserted or not in the setup 
depending on the particular measurement (see text for details). Inset: spatial mode matching on the BS.}
	\label{fig:schema}
\end{figure}
In our case we applied this procedure by considering as output qubit either the
polarization or momentum of photon $B$, demonstrating the QC equivalence of the two degrees of freedom.
This corresponds to
respectively choose the order a) or b) for the physical qubits.

The measurement apparatus is sketched in fig. \ref{fig:schema}. The $\bf k$ modes corresponding to photons A or B, are 
respectively matched on the up and down side of a common symmetric beam splitter (BS) (see inset), which
can be also finely moved in the vertical direction such that one or both photons don't pass through it.
 
Polarization analyses are performed by standard $\pi$ tomographic apparatus 
($\lambda/4$, $\lambda/2$ and polarizing beam splitter PBS),
while two HW oriented at 22.5$^\circ$ 
($H_a$ and $H_b$ in fig. \ref{fig:schema}) are inserted in photon A modes. 
They will be used together with the $\lambda/4$'s in order to transform the
$\{\ket{\varphi_+}_{\pi_A},\ket{\varphi_-}_{\pi_A}\}$ states into linearly polarized states.
Two thin glass plates before the BS allow to set the basis of momentum measurement for each photon.
\begin{figure}
\begin{center}
\includegraphics[scale=0.4]{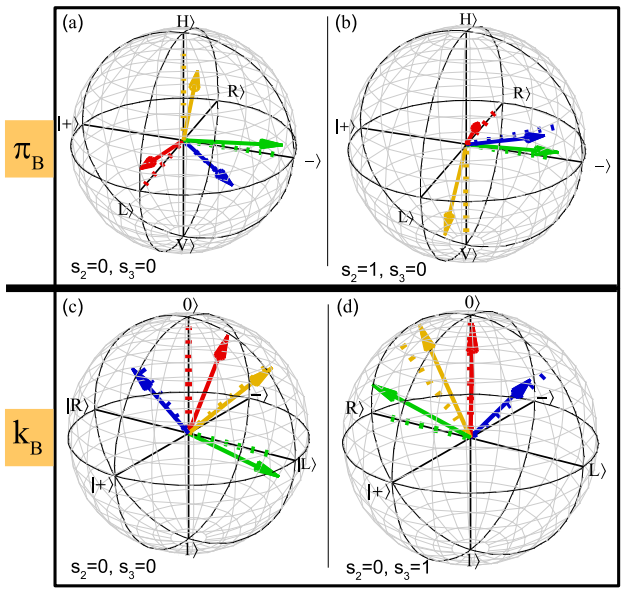}
\end{center}
\caption{Polarization ($\pi_B$) and momentum ($\bf k_B$) output Bloch vectors of single qubit rotations. 
The experimental results (arrows) are shown with their projections on theoretical directions (dashed lines). 
Arrow colours correspond to different values of $\alpha$ and $\beta$ (see table \ref{table:rotation}).  
			}
			\label{fig:bloch}
\end{figure}
\begin{table}
	\begin{ruledtabular}
				\centering
		\begin{tabular}{c|c|c|c|c|c}
			&$\alpha$ & $\beta$ & F $(s_2=s_3=0)$ & F $(s_2=1,s_3=0)$ & Colour
			\\
			\cline{2-6}					
			\cline{2-6}					
			&$0$      &  $\pi/2$    & $0.908\pm0.006$ & $0.928\pm0.013$ & orange
			\\
			\cline{2-6}					
			$\pi_B$ &$-\pi/2$      &  $0$    & $0.942\pm0.004$ & $0.902\pm0.007$ & red
			\\
			\cline{2-6}					
			&$-\pi/2$      &  $\pi/2$   & $0.913\pm0.005$ & $0.904\pm0.010$ &green
			\\
			\cline{2-6}					
			&$-\pi/2$      &  $-\pi/4$   & $0.942\pm0.006$ & $0.955\pm0.012$ &blue 
			\end{tabular}
			\begin{tabular}{c|c|c|c|c}
			&$\alpha (\beta=0)$ & F $(s_2=s_3=0)$ & F $(s_2=0,s_3=1)$ & Colour
			\\
			\cline{2-5}					
			\cline{2-5}				
			&$0$ & $0.961\pm0.003$ & $0.971\pm0.003$ & red
			\\
			\cline{2-5}					
			$\bf k_B$&$\pi/2$ & $0.879\pm0.006$ & $0.895\pm0.005$ &green	
			\\
			\cline{2-5}
			&$\pi/4$ & $0.998\pm0.005$ & $0.961\pm0.006$ & orange
			\\	
			\cline{2-5}
			&$-\pi/4$ & $0.833\pm0.007$ & $0.956\pm0.006$ & blue
\end{tabular}
\end{ruledtabular}
		\caption{Polarization ($\pi_B$) and momentum ($\bf k_B$) 
		experimental fidelities (F) of single qubit rotation output states for different values of $\alpha$ and $\beta$. 
		Each datum is obtained by the measurements of the different Stokes parameters, each one lasting 10 sec. 
		Colours in the last column correspond to those shown in fig. \ref{fig:bloch}.}
\label{table:rotation}
\end{table}

Let's consider ordering a).  
The output state, encoded in the polarization of photon B, can be written in the laboratory basis as
\eq\label{out}
\ket{\chi_{out}}_{\pi_B}={(\sigma_z)}^{s_3}{(\sigma_x)}^{s_2}HR_x(\beta)R_z(\alpha)\ket{\chi_{in}}\,,
\fine
where the $H$ gate derives from the change between the computational and laboratory basis.
This also implies that the actual measurement bases are $\ket{\pm}_{\bf k_B}$ for
the momentum of photon B and $\ket{\alpha_{\mp}}_{\bf k_A}$ for the momentum of photon A.
The measurement basis on the third qubit ($\pi_A$) 
depends, according to the one-way model, on the results of the measurement on the second qubit ($\bf k_A$).
Note that in our scheme this simply corresponds to measure it in the bases $\ket{\beta_\pm}_{\pi_A}$ or
$\ket{-\beta_\pm}_{\pi_A}$ depending on the BS output mode (i.e. $s_2=0$ or $s_2=1$).
This is a direct consequence of the possibility to encode two qubits ($\bf k_A$ and $\pi_A$) in the same photon.
As a consequence, differently from the case of four-photon cluster states,
in this case active feed-forward measurements (e.g. adopting Pockels cells) are not required, while Pauli
errors corrections are in any case necessary for deterministic QC.

In fig. \ref{fig:bloch}(a) the results obtained in the case $s_2=s_3=0$ (i.e. when the computation
proceeds without errors) with $\ket{\chi_{in}}=\ket{+}$ are given. 
We report on the Bloch sphere the experimental output qubits and their projections on the theoretical state 
$HR_x(\beta)R_z(\alpha)\ket{+}$ for different values of $\alpha$ and $\beta$. 
The corresponding fidelities are given in table \ref{table:rotation}.
We also performed the tomographic analysis on the output qubit $\pi_B$ for all the possible combinations of $s_2$ and $s_3$ and for input
qubit $\ket{\chi_{in}}=\ket{\pm}$. In all the cases we obtained an average value of fidelity $F\approx 0.9$.
As an example, we show in fig \ref{fig:bloch}(b) the case $s_2=1$, $s_3=0$ ($\ket{\chi_{in}}=\ket{+}$).
The high values of the fidelity obtained in these measurements 
represent the necessary condition to implement efficient active feed-forward corrections.

By considering ordering b) 
the same computation can be performed with output state $\ket{\chi_{out}}_{\bf k_B}$ 
encoded in the linear momentum of photon B, whose explicit expression in the laboratory basis is
\eq\label{out-momentum}
\ket{\chi_{out}}_{\bf k_B}={(\sigma_z)}^{s_3}{(\sigma_x)}^{s_2}\sigma_zHR_x(\beta)R_z(\alpha)\ket{\chi_{in}}\,.
\fine
By the apparatus of fig. \ref{fig:schema} we measured $\ket{\chi_{out}}_{\bf k_B}$ 
by 
choosing different values of $\alpha$ (corresponding in the laboratory to the polarization basis $\ket{\alpha_{\mp}}_{\pi_A}$)
and $\beta=0$ (corresponding in the laboratory to the momentum basis $\ket{-\beta_{\pm}}_{\bf k_A}$),
while the first qubit ($\pi_B$) was always measured in the basis $\ket{\pm}_{\pi_B}$.
In this case only $H_a$ and $H_b$ Hadamard gates are inserted.
The complete single qubit tomography on $\bf k_B$ requires the measurement 
of the $\sigma_x,\sigma_y$ operators, by proper setting of phase $\phi_B$, and
$\sigma_z$, performed by removing the BS on photon B.

Fig. \ref{fig:bloch}(c) shows the results obtained for different output qubits, for $s_2=s_3=0$ and $\ket{\chi_{in}}=\ket{+}$. 
The corresponding fidelities are given table \ref{table:rotation}.  
Also in this case we performed the $\bf k_B$ tomographic analysis for all the possible values of $s_2$ and $s_3$ and of the input qubit,
obtaining in average $F>0.9$. 
The case $s_2=0$, $s_3=1$ is shown in fig. \ref{fig:bloch}(d) with fidelities given in table \ref{table:rotation}.

{\bf C-NOT gate for equatorial qubits}. 
Nontrivial two-qubit operations, such as the C-NOT gate, can be realized by the four-qubit horseshoe (180$^\circ$ rotated) 
cluster state (see fig. \ref{fig:cluster}b)),
whose explicit expression is equal to $\cluster$.
By simultaneously measuring qubits 1 and 4, it's possible to implement the logical circuit shown
in fig. \ref{fig:cluster}(b). In the computational basis the input state is $\ket{+}_c\otimes\ket{+}_t$ (c=control, t=target),
while the output state, encoded in qubits 2 (control) and 3 (target),
is $\ket{\Psi_{out}}=H_tC\text{-}NOT(\mathcal O\ket{+}_c\otimes R_z(\alpha)\ket{+}_t)$
(for $s_1=s_4=0$).
In the above expression we have $\mathcal O=\openone$ ($\mathcal O=H$) when qubit 1 is measured in the basis $\{\ket0_1,\ket1_1\}$
($\ket{\pm}_1$). Qubit 4 is measured in the basis $\ket{\alpha_{\pm}}_4$.
It is worth noting that this circuit realizes the C-NOT gate (up to the Hadamard $H_t$)
for arbitrary equatorial target qubit and control qubit $\ket0,\ket1$ or $\ket{\pm}$ depending on the
measurement basis of qubit 1.
\begin{table}[t]
		\begin{ruledtabular}
						\centering
		\begin{tabular}{c|c|c|c|c}
			$\quad\mathcal O\quad$ & $\alpha$ & $\text{Control output}$ &  $F(s_4=0)$ & $F(s_4=1)$
			\\
			\hline\hline			
			             &   $\pi/2$ & $s_1=0\rightarrow\ket1_c$  & $0.965\pm0.004$ & $0.975\pm0.004$
			\\
			\cline{3-5}
			$H$          &    &$s_1=1\rightarrow\ket0_c$  & $0.972\pm0.004$ & $0.973\pm0.004$
			\\
			\cline{2-5}
			& $\pi/4$  &$s_1=0\rightarrow\ket1_c$  & $0.995\pm0.008$ & $0.902\pm0.012$
			\\
			\cline{3-5} 
      &			&$s_1=1\rightarrow\ket0_c$   & $0.946\pm0.010$ & $0.945\pm0.009$
			\end{tabular}
			\begin{tabular}{c|c|c|c|c}
  		$\ \mathcal O\ $ & $\alpha$ & $\text{Control output}$ &  $F(s_1=s_4=0)$ & $F(s_1=0,s_4=1)$
	 		\\
	 		\hline\hline
  		&  $\pi/2$  &$\ket0_c\equiv\ket \el_{\bf k_B}$  & $0.932\pm0.004$ & $0.959\pm0.003$
			\\
			\cline{3-5}
			$\openone$&&$\ket1_c=\ket r_{\bf k_B}$    & $0.941\pm0.005$ & $0.940\pm0.005$
			\\
			\cline{2-5} 
			&  $\pi/4$  &$\ket0_c=\ket \el_{\bf k_B}$  			& $0.919\pm0.007$ & $0.932\pm0.007$
			\\
			\cline{3-5}
			& &$\ket1_c=\ket r_{\bf k_B}$    	& $0.878\pm0.009$ & $0.959\pm0.006$
		\end{tabular}
		\end{ruledtabular}
		\caption{Experimental fidelity (F) of C-NOT gate output target qubit for different value of $\alpha$ and $\mathcal O$.} 
		\label{table:c-not}
\end{table}

The experimental realization of this gate is performed by adopting ordering c).
In this case the control output qubit is encoded in the momentum $\bf k_B$, while the target
output is encoded in the polarization $\pi_B$. 
In the actual experiment we inserted $H_c$ and $H_d$ on photon B to compensate $H_t$,
(while $H_a$ and $H_b$ were removed, see fig. \ref{fig:schema}).
The output state in the laboratory basis is then
\eq
\ket{\Psi_{out}}=(\Sigma)^{s_4}\sigma^{(c)}_x
C\text{-}NOT(\mathcal O\sigma^{s_1}_z\ket{+}_c\otimes R_z(\alpha)\ket{+}_t)\,,
\fine
where all the possible measurement outcomes of qubits 1 and 4 are considered. The Pauli errors are
$\Sigma=\sigma^{(c)}_z\sigma^{(t)}_z$, while the matrix $\sigma^{(c)}_x$ is due to the changing
between computational and laboratory basis.
Table \ref{table:c-not} shows the experimental fidelities of the target qubit
corresponding to the measurement of the output control qubit in the basis $\{\ket0,\ket1\}$.

The results of our experiment indicate that a two-photon four-qubit cluster state, 
which realizes the full entanglement of two photons through two degrees of freedom (in our case polarization and linear momentum),
represents a basic resource for one-way QC. 
Indeed, any kind of single qubit rotations on the Bloch sphere can be realized by these states with high fidelity. 
These transformations were performed at average repetition rates of $\sim$1~kHz by equivalently using polarization or momentum as output qubit.
The two photon approach allows to implement this algorithm without the need of active devices for 
feed-forward measurements, as demonstrated in the present work in the case of polarization output qubit.
One-way QC requires highly efficient active feed-forward corrections at the end of the process \cite{07-pre-hig}. 
The high fidelity of the output states obtained in this work, even in presence of Pauli errors,
is a necessary condition for deterministic QC.
By using the same cluster states, we also realized a C-NOT gate for target qubits located in the equatorial plane of the Bloch sphere.
Other two qubit algorithms, such as Grover's algorithm and C-Phase gate, were also implemented by us with the same source.
Results will be reported elsewere.

More complex algorithms could be realized by increasing the number of entangled qubits in the state.
For instance, six qubits are necessary to implement a C-NOT gate operating over the entire Bloch sphere. 
In our scheme more qubits could be entangled by using different degrees of freedom, such as
time-energy or exploiting the continuous k-mode emission within the SPDC cone of a type I crystal,
and/or increasing the number of photons. For example eight-qubit four-photon cluster state
cloud be generated by linking together two $\ket{C_4}$ states by a proper CP gate.
These different approaches are at the moment under investigation.

\begin{acknowledgments}
Thanks are due to Fabio Sciarrino for useful discussions and Marco Barbieri for
his contribution in planning the experiment. This work was supported by the PRIN 2005 of MIUR (Italy).
\end{acknowledgments}

{\it Note added-} While we were completing the experiment
another work concerning one-way QC with two-photon four-qubit cluster states appeared \cite{07-che-exp}.


\end{document}